\documentclass[aip, floatfix, jcp, reprint, longbibliography]{revtex4-1}

\usepackage{enumitem}
\usepackage{graphicx}          
\usepackage{amsmath}    
\usepackage{amssymb}    
\usepackage{amsthm}
\usepackage{graphicx}   
\usepackage{float}
\usepackage{hyperref}   
\usepackage[capitalise]{cleveref}   
\usepackage{appendix}
\usepackage{color, soul}
\usepackage{bm}

\newcommand{\fu}{Department of Mathematics and Computer Science, Freie Universit{\"a}t Berlin, Arnimallee 6, 14195 Berlin, Germany}
\newcommand{\stanford}{Department of Chemistry, Stanford University, 333 Campus Drive, Stanford, California 94305, USA}
\newcommand{\rice}{Department of Chemistry, Rice University, 6100 Main St., Houston, Texas 77005, USA}

\begin{document}

\title{Variational Selection of Features for Molecular Kinetics}

\author{Martin K. Scherer}
\thanks{M.K.S.~and B.E.H.~contributed equally to this work.}
\affiliation{\fu}
\author{Brooke E. Husic}
\thanks{M.K.S.~and B.E.H.~contributed equally to this work.}
\affiliation{\fu}
\affiliation{\stanford}
\author{Moritz Hoffmann}
\affiliation{\fu}
\author{Fabian Paul}
\affiliation{\fu}
\affiliation{Department of Biochemistry and Molecular Biology, Gordon Center for Integrative Science, The University of Chicago, Chicago, IL 60637, USA}
\author{Hao Wu}
\affiliation{\fu}
\affiliation{School of Mathematical Sciences, Tongji University, Shanghai, 200092, P.R. China}
\author{Frank No{\'e}}
\email{frank.noe@fu-berlin.de}
\affiliation{\fu}
\affiliation{\rice}


\begin{abstract}
The modeling of atomistic biomolecular simulations using kinetic models such as Markov state models (MSMs) has had many notable algorithmic advances in recent years. The variational principle has opened the door for a nearly fully automated toolkit for selecting models that predict the long-timescale kinetics from molecular dynamics simulations. However, one yet-unoptimized step of the pipeline involves choosing the features, or collective variables, from which the model should be constructed. In order to build intuitive models, these collective variables are often sought to be interpretable and familiar features, such as torsional angles or contact distances in a protein structure. However, previous approaches for evaluating the chosen features rely on constructing a full MSM, which in turn requires additional hyperparameters to be chosen, and hence leads to a computationally expensive framework. Here, we present a method to optimize the feature choice directly, without requiring the construction of the final kinetic model. We demonstrate our rigorous preprocessing algorithm on a canonical set of twelve fast-folding protein simulations, and show that our procedure leads to more efficient model selection. 
\end{abstract}

\maketitle

\section{Introduction}

The first step in analyzing the states, equilibrium behavior, or kinetics of complex molecules based on molecular dynamics (MD) simulations is typically the choice of a suitable set of features describing the atomic configurations. This choice is particularly important when the goal is to compute kinetic quantities, such as transition rates or committor probabilities, as these quantities are sensitive to resolving the transitions between the long-lived (metastable) states~\cite{hanggi1983memory,shalloway1996macrostates,du1998transition,rohrdanz2013discovering,prinz2014spectral,noe2017collective}. Typically, these transformations input the raw Cartesian coordinates produced from MD simulation and output a new set of coordinates that is translation- and rotation-invariant. In proteins, biologically-motivated choices include the backbone dihedral angles (torsions), or the pairwise distances between all amino acid residues taken between $\alpha$-carbons or the closest pairs of (heavy) atoms. Many coordinate sets for proteins are imaginable, such as further transformations of the aforementioned contact distances, the solvent-accessible surface area (SASA) of protein residues or other groups, or sidechain torsion angles.

A number of kinetic analysis frameworks critically depend on the set of features used as input, in particular Markov state models (MSMs)~\cite{schutte1999direct,swope2004adescribing,noe2007hierarchical,chodera2007automatic,singhal2004using,prinz2011markov,husic2018markov}, master equation models~\cite{sriraman2005coarse,buchete2008coarse}, diffusion maps~\cite{rohrdanz2011determination}, and methods to select optimal reaction coordinates~\cite{peters2006using,e2006towards}.
The analysis framework employed affects what is a meaningful definition of optimality that can be used to select input features. 

Here we discuss optimal feature selection in particular with MSMs in mind.
In the first generation of MSM methodology, whose aim was to construct a Markov chain between a few metastable states that partition state space~\cite{schutte1999direct,swope2004adescribing,noe2007hierarchical,chodera2007automatic}, it was already noted that the accuracy with which MSMs could make long-timescale predictions depended critically on the choice of features. Example featurization methods included torsion angles~\cite{hummer2005position,chodera2006long,noe2007hierarchical,kube2007coarse}, principal components in torsion or Cartesian coordinates~\cite{altis2007dihedral,noe2009constructing}, contact pairs~\cite{jayachandran2006parallelized,yang2008src}, as well as other transformations involving such attributes as secondary structure~\cite{buchete2008coarse,muff2008kinetic}. From these representations, clustering methods can be used to determine the MSM states, which can be subsequently checked for adherence to the Markovian approximation.
The second generation of MSM methodology is characterized by the finding that the predictive power of an MSM depends upon its states being chosen such that a good discretization of the Markov operator eigenfunctions is obtained~\cite{sarich2010approximation,prinz2011markov}. These eigenfunctions are collective variables that indicate the rare-event processes connecting the metastable states, and the task of discretizing them well translates into the task of using input features that allow to resolve them.

At this time, mainly visual diagnostic tools were available to assess the predictive performance of MSMs, such as the implied timescales test~\cite{swope2004adescribing} and the Chapman-Kolmogorov test~\cite{prinz2011markov}. The choice of the hyperparameters of MSMs, such as number of states and the set of input features, remained a trial-and-error procedure. This changed in 2013, when No{\'e} and N{\"u}ske presented a variational principle that quantifies how well a given set of coordinates, features or a given MSM resolve the Markov operator eigenfunctions, and thus the slowest processes~\cite{noe2013variational,nuske2014variational}. This variational approach to conformational dynamics (VAC) has been highly developed in the past five years:~time-lagged independent component analysis (TICA), an algorithm devised in machine learning~\cite{molgedey1994separation}, has been shown to be the optimal linear approximator to the Markov operator eigenfunctions~\cite{perez2013identification}; an embedding of the eigenfunctions approximated by the VAC or TICA into a kinetic map has been proposed, in which distances are related to transition times~\cite{noe2015kinetic,noe2016commute};and hierarchical~\cite{perez2016hierarchical} and kernel-based estimators~\cite{schwantes2015modeling,harmeling2003kernel} have been developed, as well as methods to estimate VAC/TICA from short off-equilibrium trajectories~\cite{wu2017variational}. Recently, the VAC has been generalized to the variational approach for Markov processes (VAMP), which can accommodate nonreversible dynamics~\cite{wu2017vamp}.

A key insight is that the variational principle defines a score which can be used with standard machine learning approaches for hyperparameter selection of MSMs. This was pioneered in Ref.~\onlinecite{mcgibbon2015variational}, which shows that using the VAC with cross-validation is a tool for selecting the statistically optimal number of MSM states. Using a VAC-derived kinetic variance as a score, optimal feature selection was discussed in Ref.~\onlinecite{scherer2015pyemma}. The entire MSM pipeline was subject to VAC-optimization in Ref.~\onlinecite{husic2016optimized}, revealing general trends in what makes a good MSM for fast protein folding. The VAMP has been used for hyperparameter optimization~\cite{wu2017vamp}, and in order to define the loss functions for VAMPnets, a deep learning method to infer MSMs from data~\cite{mardt2018vampnets}.

The construction of an MSM involves (a)~choosing an appropriate transformation to collective variables, referred to as features; (b)~optionally performing a basis set transformation using TICA (or, alternatively, stopping here and using the TICA result as the kinetic model); (c)~decomposing the transformed trajectories into states; and (d)~approximating a Markovian transition matrix from the state decomposition.
Step (a)~is difficult to automate using current methods.
In contrast, accounting for the relatively established use of just a few clustering algorithms for the state decomposition~\cite{scherer2015pyemma,husic2016optimized}, steps~(b) and~(c) have largely been automated using the VAC---and, more recently, the VAMP.
To optimize~(b) and~(c), it is straightforward to select an arbitrary number of states, construct a few hundred cross-validated MSMs, and identify which number of states achieves the highest VAMP score.
While this can in principle also be done for collective variable transformations (a) as well, repeated construction of the entire MSM pipeline for a variety of input features becomes computationally extremely demanding. A complete search of a hyperparameter space (features, TICA dimension, clustering method and number of clusters, etc.)~is clearly unfeasible. Hence a variationally optimal method for feature selection that does not require going through the additional steps and choices of building an MSM would be very useful.

Motivated by these difficulties, in this paper we describe an approach that introduces a theoretically rigorous method for the choice of input features and an accompanying algorithm which enables the researcher to quantify this choice. This could in principle also be used to automate feature selection. We study this approach by applying the method to a canonical dataset of twelve fast-folding proteins simulated near their experimental melting temperatures~\cite{kubelka2004protein, lindorff2011fast}. These systems switch between the folded and unfolded state rapidly, and each trajectory dataset contains at least 10 instances of both folding and unfolding. The dataset, which contains fast folding proteins possessing a variety of secondary structure combinations, has been frequently used in its entirety to investigate methods advances~\cite{beauchamp2012simple,dickson2013native,weber2013emergence,husic2016optimized}. After evaluating features on all twelve proteins, we build representative MSMs to illustrate an example analysis.

Associated code is available at \texttt{github.com/markovmodel/feature\_selection}. The computation of VAMP scores is implemented in the VAMP estimator method of the PyEMMA software package~\cite{scherer2015pyemma}, which can be found at \texttt{github.com/markovmodel/PyEMMA}.

\section{Theory} \label{sec:theory}

Here we summarize the necessary theory underlying VAMP-based feature selection, starting with MSMs and the VAC before continuing on to the Koopman matrix and the VAMP used in this work. The more practically inclined reader may proceed to Sec.~\ref{sec:methods}---the main point of this section is that the VAMP score of a given set of features can be computed by implementing equations~(\ref{eqn:c00}-\ref{eqn:vamp-r}). For more detailed theoretical discussions, we refer the reader to Refs.~\onlinecite{wu2017variational},~\onlinecite{wu2017vamp}, and~\onlinecite{paul2018identification}.

\subsection{The Markov state model transition matrix}

MSMs model dynamics as a coarse-grained Markov chain on sets $A_i$ that partition the state space. For this discussion, we assume a single long trajectory, although the method may be applied to a distributed set of independent trajectories.
The Markov chain is described by the conditional transition probabilities,

\begin{align}
p_{ij}(\tau) \equiv \text{Pr}(x_{t+\tau}\in A_j | x_t \in A_i),
\label{eqn:msmmaster}
\end{align}

\noindent{}where $p_{ij}(\tau)$ represents the probability that the system $x$ is in set $A_j$ at time $t+\tau$ conditioned upon it being in set $A_i$ at time $t$. These probabilities can be estimated from trajectories initiated from local equilibrium distributions, and do not require that the system is in global equilibrium~\cite{prinz2011markov}.

The conditional transition probabilities $p_{ij}(\tau)$ are gathered in a square probability transition matrix $\hat{\mathbf{P}}(\tau)$, where each entry represents the conditional probability of transitioning from the set described by row index $i$ to the set described by column index $j$ at the defined lag time $\tau$. When a sufficiently long lag time $\tau$ and adequate sets have been chosen, the dynamics can be approximated as Markovian, and the $p_{ij}(\tau)$ are independent of the history of the system. Thus, many independent trajectories obtained from distributed simulations can be threaded together through common sets (henceforth, ``states'').

Once a simulation dataset has been divided into states, the $p_{ij}(\tau)$ estimates are obtained from an analysis of observed transition counts.
The observed transition counts are converted into conditional transition probabilities such that the dynamics are reversible.
For MSM analysis, the eigendecomposition of the reversible transition matrix $\hat{\mathbf{P}}(\tau)$ contains information about the thermodynamics and kinetics of the modeled system. Its stationary distribution is given by the eigenvector corresponding to the unique maximum eigenvalue $\lambda_1 = 1$. The remaining eigenvalues, which are restricted to the interval $|\lambda_i| < 1$ for $i\geq 2$, correspond to dynamical processes within the system. Timescales of these processes are defined as a function of the eigenvalue and the MSM lag time: 

\begin{align}
t_i\left[ \hat{\mathbf{P}}(\tau)  \right] \equiv \frac{-\tau}{\log | \lambda_i(\tau)|},
\label{eqn:implied_timescales}
\end{align}

\noindent{}where the absolute values are used by convention to avoid the imaginary timescales resulting from projection of the system dynamics.

To test whether the Markovian assumption is appropriate for an MSM approximated at a given lag time, the timescales can be plotted as a function of increasing lag time to observe if they have converged to an approximately constant value at the lag time of the estimator; this is referred to as validating the model's implied timescales~\cite{swope2004adescribing}. Evaluation of the implied timescales is special case of a more general validation tool, the Chapman-Kolmogorov test, which evaluates the appropriateness of the Markovian assumption according to adherence to the property\cite{prinz2011markov},

\begin{align}
[\hat{\mathbf{P}}(\tau)]^k \approx \hat{\mathbf{P}}(k\tau). \label{eqn:ck}
\end{align}


\subsection{The variational approach for conformational dynamics (VAC)}

The eigenvalues and eigenvectors of the transition probability matrix correspond to the stationary and dynamical processes in the system, and these quantities can be used to interpret MD simulation data once states have been determined and the MSM is constructed~\cite{prinz2011markov,husic2018markov}. 
The discrete eigenvectors are in fact approximations to continuous eigenfunctions, and the transition probability matrix is a finite-dimensional approximation to an infinite-dimensional continuous linear operator called the transfer operator, which describes the system dynamics~\cite{schutte1999direct,prinz2011markov,schutte2015critical}.
In an MSM the system is described by a disjoint set of discrete states, which can be represented by a basis set of indicator functions $\{\xi_i(x)\}$, i.e.,

\begin{align}
\xi_i(x) \equiv
  \begin{cases}
    1, &\text{if }x\in A_i\\
    0, &\text{otherwise}.
  \end{cases}
  \label{eqn:indicator}
\end{align}

Importantly, the VAC generalizes beyond MSMs and does not require an indicator basis set~\cite{noe2013variational,nuske2014variational}. Rather, the VAC can be applied to an \emph{arbitrary} basis set $\{\chi_i(x)\}$.  Given a basis set, we then transform all observed $x$ in the dataset and estimate two matrices from the transformed data:~$\mathbf{C}(0)$, the covariance matrix, and $\mathbf{C}(\tau)$, the cross-covariance matrix between the data and the time-lagged data,~\footnote{For more details on this estimation, see Ref.~\onlinecite{wu2017variational}, \S II C.}

\begin{align}
\mathbf{C}(0)&=\mathbb{E}_{\mu_0} [\bm{\chi}(x_t)\bm{\chi}(x_t)^\top], \label{eqn:czero} \\
\mathbf{C}(\tau)&=\mathbb{E}_{\mu_0} [\bm{\chi}(x_t)\bm{\chi}(x_{t+\tau})^\top], \label{eqn:ctau}
\end{align}

\noindent{}where the $\bm{\chi}$ are matrices containing the feature vectors $\{ \chi_i(x)\}$, and the subscript $\mu_0$ on the expected value $\mathbb{E}$ indicates that $x_t$ is sampled from the stationary distribution $\mu$ on the interval $[0, T-\tau]$, for $T$ total time.

With these estimates, we can proceed to solve the generalized eigenvalue problem,

\begin{align}
\mathbf{C}(\tau)\mathbf{B}&=\mathbf{C}(0)\mathbf{B}\hat{\bm{\Lambda}},
\label{eqn:generalizedeig}
\end{align}

\noindent{}where $\hat{\bm{\Lambda}} = \text{diag}(\hat{\lambda}_1, \dots, \hat{\lambda}_m )$, and the matrix $\mathbf{B}$ provides vectors of expansion coefficients $\{b_1,\dots,b_m\}$, which can be substituted into the \textit{ansatz} to obtain the approximated eigenfunctions $\{ f_i \}$~\cite{noe2013variational,nuske2014variational,wu2017variational}:

\begin{align}
f_i(x) = \sum_{j=1}^{m}b_{ij}\chi_j(x).
\label{eqn:ansatz}
\end{align}

\noindent{}Since the \textit{ansatz}~\eqref{eqn:ansatz} is linear in the vector $b_i$, this characterizes the linear VAC.
However, it is just this last step that is linear:~the MSM eigenfunctions can be arbitrarily nonlinear based on the choice of the basis set $\{ \chi_i(x) \}$.
The expansion coefficients $\{ b_i \}$ are chosen such that the $\{f_i (x)\}$ maximize the Rayleigh trace,

\begin{align}
R_m = &\sum_{i=1}^m\mathbb{E}_{\mu_0}[f_i(x_t)f_i(x_{t+\tau})],\label{eqn:gmrq} \\
&\text{ such that } \mathbb{E}_{\mu_0}[f_i(x_t)f_j(x_t)]=\delta_{ij}, \nonumber
\end{align}

\noindent{}where $\delta_{ij}$ is the Kronecker delta. The definition of the score $R_m$ turns MSM estimation into a machine learning problem where tools such as cross-validation can be used to determine hyperparameters~\cite{mcgibbon2015variational}.

\subsection{Estimating Koopman matrix and VAMP score}

In Ref.~\onlinecite{wu2017vamp}, a new variational approach is introduced that does not require the operator it approximates to be reversible, nor does it require simulation data that are in equilibrium.
The operator is the Koopman operator, which is approximated by the Koopman matrix.
The corresponding generalized master equation is given by,

\begin{align}
\mathbb{E}[\mathbf{g}(x_{t+\tau})] = \hat{\mathbf{K}}^\top(\tau)\mathbb{E}[\mathbf{f}(x_t)],
\label{eqn:koopmanmaster}
\end{align}

\noindent{}where $\mathbf{f}$ and $\mathbf{g}$ are matrices storing the feature transformations $\{ f_i \}$ and $\{ g_i \}$, respectively.

\citet{wu2017vamp} show that optimal choices for $\mathbf{f}$ and $\mathbf{g}$ can be determined using the singular value decomposition (SVD) of the Koopman matrix and setting $\mathbf{f}$ and $\mathbf{g}$ to its top left and right singular functions, respectively.\footnote{Ref.~\onlinecite{wu2017vamp}, Thm.~1.}
For an MSM, $f$ and $g$ are basis sets of indicator functions as defined in Eqn.~\eqref{eqn:indicator}, and Eqn.~\eqref{eqn:koopmanmaster} is equivalent to Eqn.~\eqref{eqn:msmmaster}.\footnote{To see the equivalence, as an intermediate step write $\mathbb{E}_{t+\tau}(\xi_i) = p_{ij}(\tau) \mathbb{E}_t(\xi_j)$.}

A new variational principle---the VAMP---can then be applied to approximate the singular functions by maximizing the Rayleigh trace as in Eqn.~\eqref{eqn:gmrq} for the $m$ dominant singular values, except without requiring $\mathbf{f}=\mathbf{g}$,

\begin{align} \label{eqn:vamp}
R^\prime_m = &\sum_{i=1}^m\mathbb{E}_{\rho_0}[f_i(x_t)g_i(x_{t+\tau})], \\
&\text{ such that } 
\begin{cases}
\mathbb{E}_{\rho_0}[f_i(x_t)f_j(x_t)]=\delta_{ij} \text{, and } \\
\mathbb{E}_{\rho_1}[g_i(x_t)g_j(x_t)]=\delta_{ij},
\end{cases} \nonumber
\end{align}

\noindent{}where $\mathbb{E}_{\rho_0}$ and $\mathbb{E}_{\rho_1}$ perform the expected value over the starting points of the time windows $[0,T-\tau]$ and $[\tau,T]$, respectively (which are no longer required to represent stationary samples), and $T$ is the total time.

The VAMP was designed to be amenable to nonreversible processes; therefore it is useful to permit $\mathbf{f}$ and $\mathbf{g}$ to be different.
This enables an adapted description of the dynamics that is different at the beginning and end of a transition of duration $\tau$, because the system may have changed to the extent that it makes sense to adapt a new basis for the system after the lag time.
Although we typically have stationary dynamics in MD datasets, we use the VAMP because we do not need to enforce reversibility in the dynamics as in the VAC, nor do we need to perform the statistically unfavorable reweighting~\footnote{While the Koopman reweighting estimator introduced in Ref.~\onlinecite{wu2017variational} removes bias, it has a relatively large variance.} described in Ref.~\onlinecite{wu2017variational}.
For the VAMP, we require the following three covariance matrices:

\begin{align}
\mathbf{C}_{00} &\equiv \mathbb{E}_{\rho_0}[\bm{\chi}(x_t)\bm{\chi}(x_t)^\top], \label{eqn:c00}\\
\mathbf{C}_{01} &\equiv \mathbb{E}_{\rho_0}[\bm{\chi}(x_t)\bm{\chi}(x_{t+\tau})^\top], \label{eqn:c01}\\
\mathbf{C}_{11} &\equiv \mathbb{E}_{\rho_1}[\bm{\chi}(x_{t})\bm{\chi}(x_{t})^\top].\label{eqn:c11}
\end{align}

In order to obtain the singular vectors of $\hat{\mathbf{K}}$, we instead must determine the singular vectors of a different matrix $\bar{\mathbf{K}}$, which represents propagation in a whitened basis set with all correlation between features removed, i.e.,

\begin{align}
\bar{\bm{\chi}}_{\rho_0} &\equiv \mathbf{C}_{00}^{-\frac{1}{2}}\bm{\chi} \\
\bar{\bm{\chi}}_{\rho_1} &\equiv \mathbf{C}_{11}^{-\frac{1}{2}}\bm{\chi},
\end{align}

\noindent{}where the subscripts $\rho_0$ and $\rho_1$ indicate whether the distribution was drawn from the data on $[0, T-\tau]$ or the time-lagged data on $[\tau, T]$, respectively.
Since the approximation $\bar{\mathbf{K}}$ is made in the whitened basis, differing feature scales do not need to be accounted for, since they would be undone in this step.
We further require $\mathbf{C}_{00}$ and $\mathbf{C}_{11}$ to be invertible.\footnote{See Ref.~\onlinecite{wu2017vamp}, Appendix F for details. When the whitening as suggested in Ref.~\onlinecite{wu2017vamp} is used, $\mathbf{C}_{00}$ and $\mathbf{C}_{11}$ become identity matrices when calculated from the whitened data.}
With these matrices, we can perform the approximation,

\begin{align}
\bar{\mathbf{K}}(\tau)=\mathbf{C}_{00}^{-\frac{1}{2}}\mathbf{C}_{01}\mathbf{C}_{11}^{-\frac{1}{2}}\approx \mathbf{U}_m\bm{\Sigma}_m\mathbf{V}_m^\top,
\label{eqn:kapprox}
\end{align}

\noindent{}where $\bm{\Sigma}_m=\text{diag}(\sigma_1,\dots,\sigma_m)$ (the first $m$ singular values), $\mathbf{U}_m$ and $\mathbf{V}_m$ contain the corresponding $m$ left and right singular vectors of the whitened Koopman matrix $\bar{\mathbf{K}}(\tau)$.

The matrices $\mathbf{U}_m$ and $\mathbf{V}_m$ are used to calculate the optimal $\mathbf{f}$ and $\mathbf{g}$ (in the original, non-whitened space) as follows:

\begin{align}
\mathbf{f} &= \mathbf{U}^\top \bm{\chi}, \\
\mathbf{g} &= \mathbf{V}^\top \bm{\chi},
\end{align}

\noindent{}where,

\begin{align}
\mathbf{U} &\equiv \bar{\mathbf{C}}_{00}^{-\frac{1}{2}}\mathbf{U}_m \text{, and} \label{eqn:u} \\
\mathbf{V} &\equiv \bar{\mathbf{C}}_{11}^{-\frac{1}{2}}\mathbf{V}_m. \label{eqn:v}
\end{align}

\noindent{}Although the last step is linear, as in MSM analyses, the choice of the basis set $\bm{\chi}$ permits arbitrary nonlinearity in the model. From $\mathbf{\Sigma}_m$ we can write the VAMP-$r$ score:

\begin{align}
\text{VAMP-}r \equiv \sum_{i=1}^m \sigma_i^r.
\label{eqn:vamp-r}
\end{align}

Previously, in the VAC, we summed the $m$ dominant eigenvalues of the MSM transition matrix to obtain the model's score, which is variationally bounded from above. In a subsequent work,~\citet{noe2015kinetic} demonstrated that the sum of the squared eigenvalues is also variationally bounded from above, and can be maximized to obtain the kinetic variance described by the approximator. In fact, any nonincreasing weight can be applied to the eigenvalues and the variational principle will still hold~\cite{gross1988rayleigh}.
Thus, we can sum the $m$ highest singular values, each raised to an exponent $r\geq 1$, to obtain the VAMP-$r$ score\cite{wu2017vamp}. VAMP-$1$ is analogous to the Rayleigh trace, and VAMP-$2$ is analogous to the kinetic variance introduced in Ref.~\onlinecite{noe2015kinetic}.\footnote{
When the stationary process is modeled as in Ref.~\onlinecite{wu2017vamp}, the score is bounded by $m$+1, where $m$ is the number of dynamical (i.e., non-stationary) processes scored.
}
\section{Methods} \label{sec:methods}

\begin{figure*}[htb!]
\centering
\includegraphics[width=\textwidth]{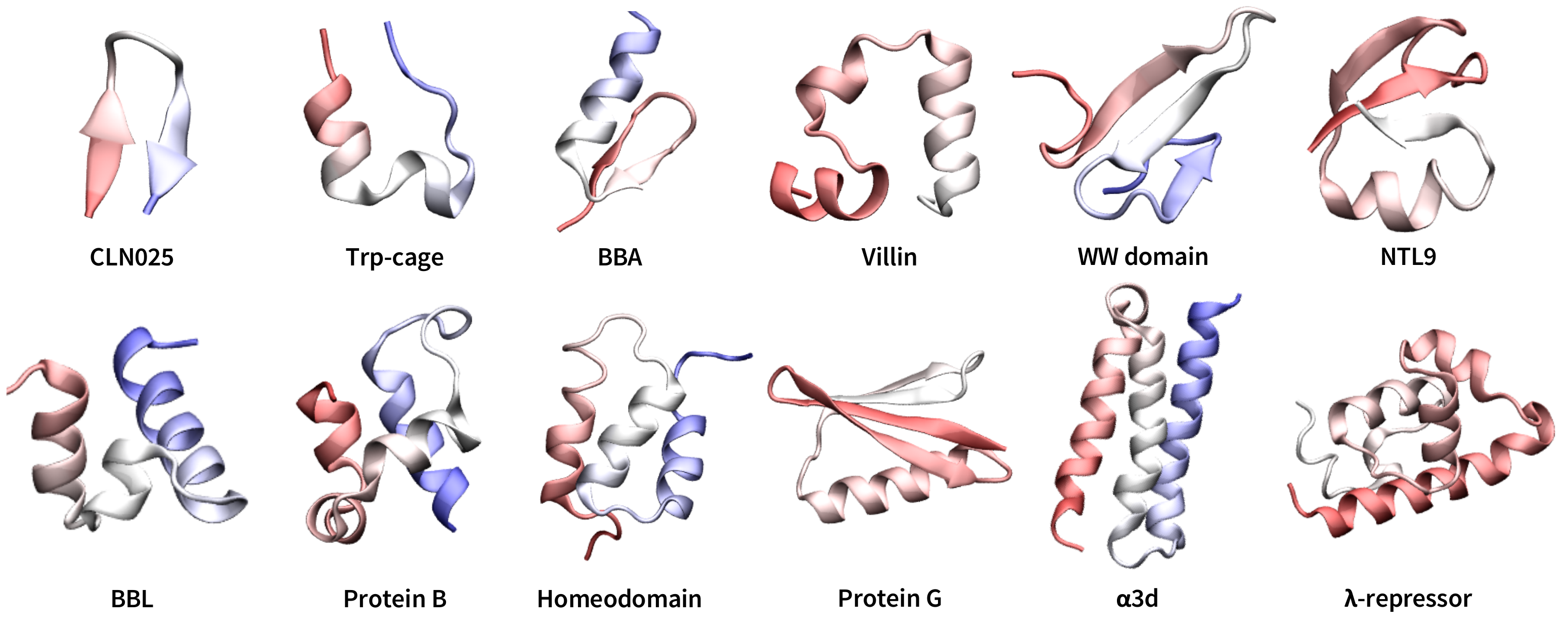}
\caption{Representative structures for twelve fast-folding proteins simulated by \citet{lindorff2011fast}. When the trajectory input file corresponded to the folded state, this structure was used. Otherwise, a folded structure was hand-selected from the trajectory dataset in order to represent a na{\"i}ve alignment choice that precedes analysis. In some cases, disordered tails of the pictured proteins are not shown. These folded states were used for the aligned Cartesian coordinate features.}
\label{fig:desres}
\end{figure*}

\begin{figure}[htb!]
\centering
\includegraphics[width=0.45\textwidth]{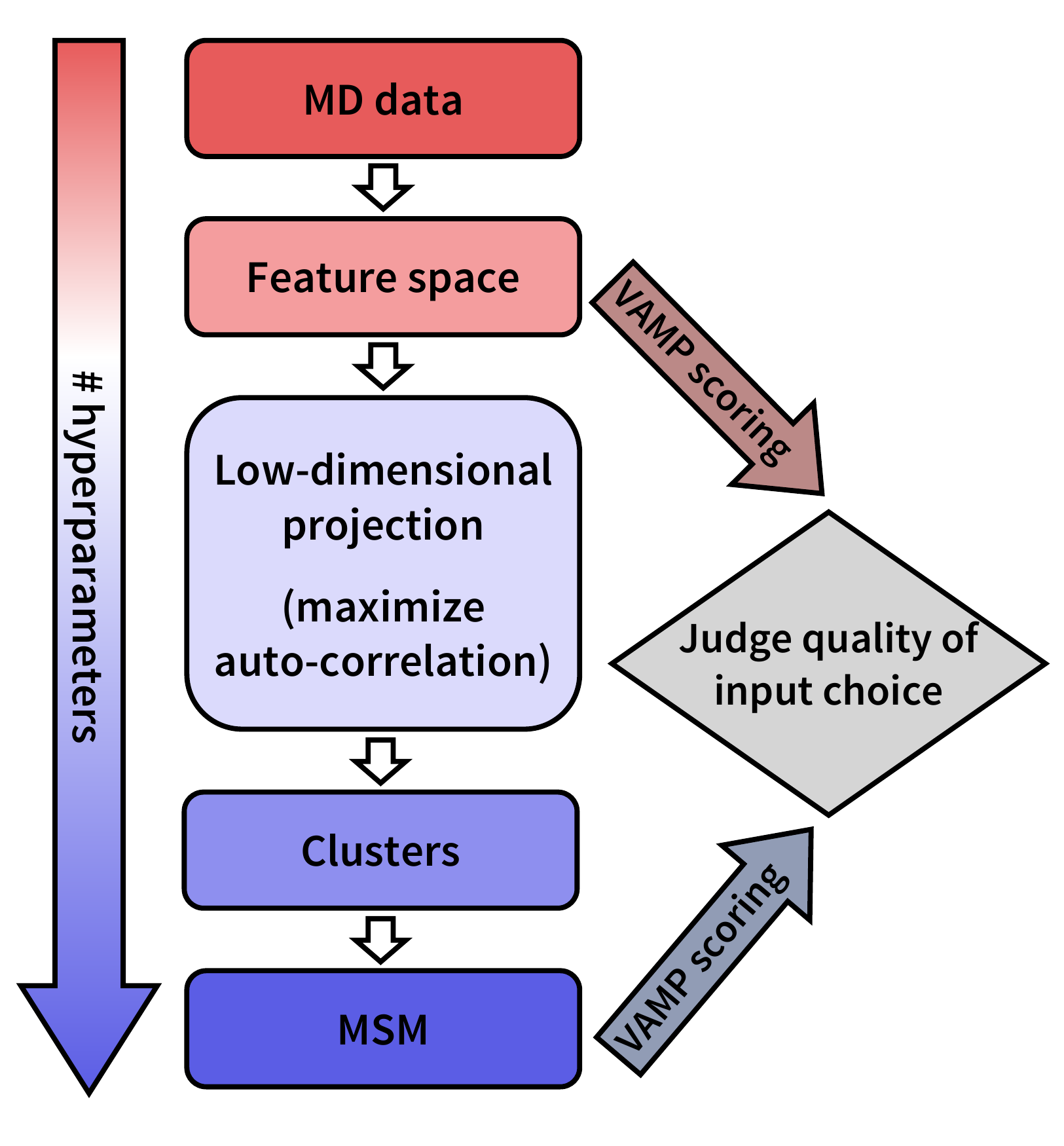}
\caption{Overview of protocol. From the MD simulations, we extract a set of features. In the feature space we compute a VAMP score to obtain a measure for the model approximation quality. Further steps towards a kinetic model involve a transformation, e.g.~TICA or VAMP projection to maximize auto-correlation, and clustering in this transformed space. In our study, we use the first five VAMP singular vectors for this step. Once discretized, the trajectories are then used to estimate a MSM. After validating the MSM, another VAMP score can be computed. All steps except the direct scoring of the feature space are established methods.}
\label{fig:procedure}
\end{figure}

\subsection{Input features}

We apply the described algorithm to a set of fast-folding protein trajectory datasets~\cite{lindorff2011fast} (Fig.~\ref{fig:desres}) in order to calculate scores for several commonly used features in Markov state modeling. We investigate the following choices of input features:

\begin{enumerate}
\item \textit{Aligned Cartesian coordinates.} We first use Cartesian coordinates, which have been aligned---i.e., translated and rotated to minimize the root mean square deviation---to the folded structure for the simulation dataset.\footnote{The folded structures were chosen visually, before analysis, to replicate a na{\"i}ve choice of reference frame. For WW domain, residues 5--30 were used for the aligned Cartesian coordinate feature, but the results are comparable for the full system.} \label{def:feat_xyz}
\item \textit{Distance-based features.} As a baseline distance feature, the closest heavy-atom contacts between each pair of residues in the protein is recorded in nanometers, excluding proximal pairs at indices $(i, i+1)$ and $(i, i+2)$. In addition to using this distance, denoted as $d \equiv \text{min}(d)$, we also apply several transformations to $d$:
	\begin{enumerate}[label=\roman*.]
    	\item $f(d) = d^{-1}$
        \item $g(d) = d^{-2}$
        \item $h(d) = \log(d)$
        \item $h^\prime(d) = e^{-d}$.
        \label{def:feat_dists}
        \end{enumerate} 
\item \textit{Contact features.} To form the contact features, we apply five different cutoff values on the residue minimum distances in order to encode a contact in a binary manner. In other words, we calculate, \label{def:feat_contact}
\begin{align*}
h_c(d) = \begin{cases}
1, & \text{if } d < c \\
0, & \text{otherwise},
\end{cases}
\end{align*}
for $c \in \{ 4, 5, 6, 8, 10\}$~\AA.
    \item \textit{Solvent accessible surface area (SASA).} SASA is computed by the Shrake-Rupley rolling ball algorithm\cite{shrake1973environment} as implemented in the MDTraj software package~\cite{mcgibbon2015mdtraj}. This algorithm estimates the surface exposed to solvent by moving a small sphere beyond the van der Waals radii of each atom of the protein. The radius was set to 1.4~\AA, and the number of grid points to model the sphere was set to 960. The SASA for each residue is calculated from the sum of the SASA results for each of its component atoms.
\item \textit{Flexible torsions.} The torsion features contain the backbone torsional (dihedral) angles $\phi$ and $\psi$ as well as all possible side chain torsions $\chi_i$ up to $\chi_5$. Each torsional angle is transformed into its sine and cosine in order to maintain periodicity.
\item \textit{Combined distance transformation and flexible torsions.} Finally, we concatenate the transformed distance feature $e^{-d}$ with the flexible torsions. \label{def:feat_combined}
\end{enumerate}

\subsection{Covariance matrix estimation and cross-validation of scores}

We will now sketch the practical algorithm for calculating the cross-validated VAMP-2 score of a feature space. As an input we have feature trajectories $\mathbf{X}_{1}, \ldots, \mathbf{X}_{n}$ with $N_{1},\ldots,N_{n}$ frames in $\mathbb{R}^{d}$. From these time series we estimate the covariance matrices $\mathbf{C}_{00}, \mathbf{C}_{01}$, and $\mathbf{C}_{11}$ (recall Eqns.~\eqref{eqn:c00}-\eqref{eqn:c11}) by an online algorithm (i.e., one that does not require all of the data to be input simultaneously into memory)\cite{chan1982updating}.

In order to compute a cross-validated VAMP score, we estimate $n_\text{cov}$ covariance matrix triplets $(\mathbf{C}_{00,k}, \mathbf{C}_{01,k}, \mathbf{C}_{11,k})$. To do so, the time series is split into temporally subsequent pairs of overlapping blocks

\begin{align}
\left(\mathbf{B}_j, \mathbf{B}_j^\prime\right) = \left((\mathbf{X}_\ell)_{i=t_j}^{t_j+2\tau},
(\mathbf{X}_\ell)_{i=t_j+\tau}^{t_j+3\tau} \right),
\label{eq:sliding_window}
\end{align}

\noindent{}where $\tau$ denotes the lag time in each trajectory $\ell$, $t$ is the current position in time, and $(\mathbf{X}_{\ell})_{i=a}^b$ the frames between $t=a$ inclusively and $t=b$ exclusively. 

\begin{figure*}[ht!]
\centering
\includegraphics[width=1\textwidth]{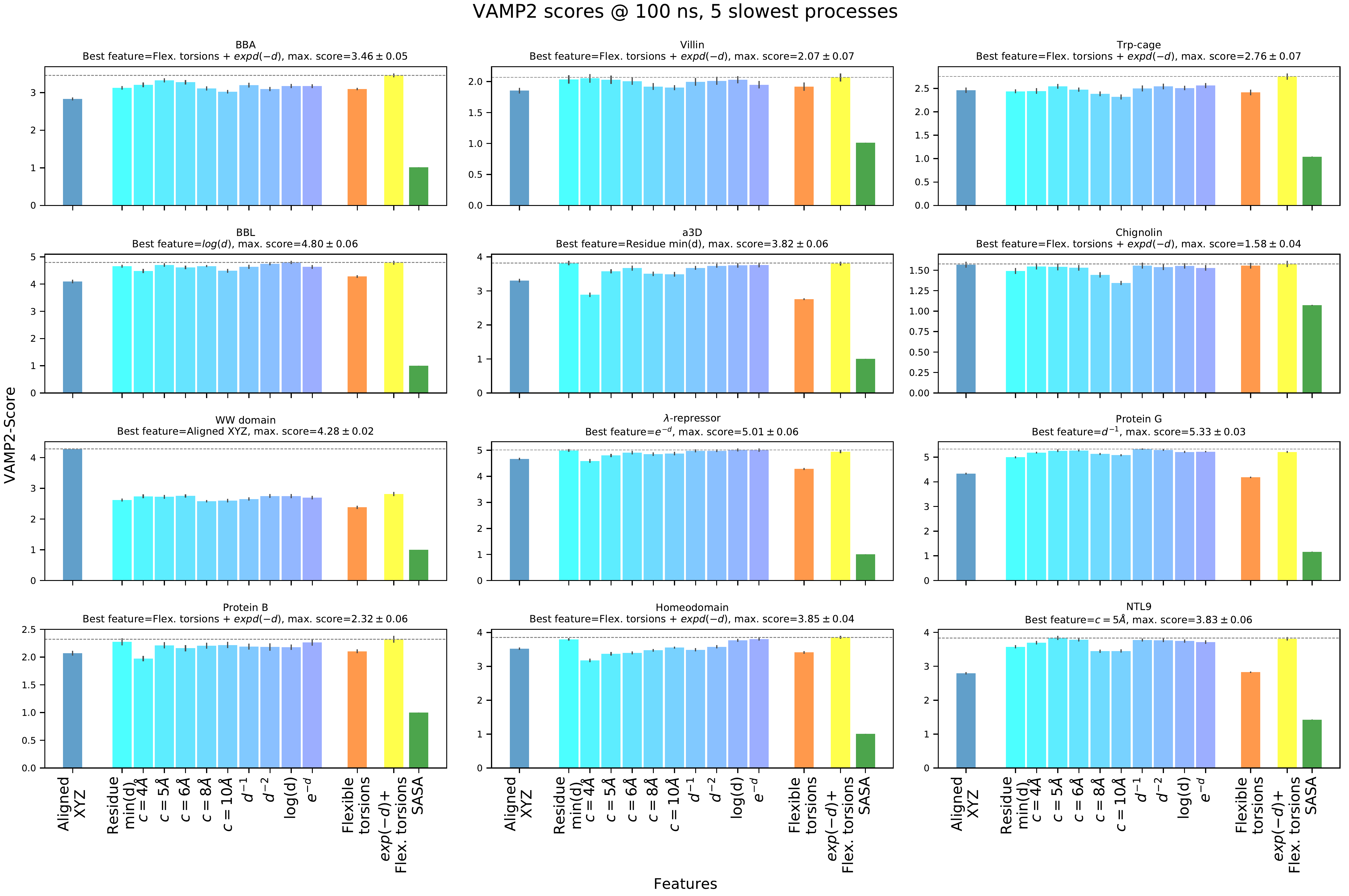}
\caption[width=1\textwidth]{VAMP-2 scores of the five slowest processes for all test systems and all defined features at a lag time of $100$ ns. Error bars represent standard errors across 50 cross-validation splits.}
\label{fig:vamp2_all_sys_k5}
\end{figure*}

For each pair of blocks $(\mathbf{B}_{i},\mathbf{B}_{i}^{\prime})$ as yielded by the sliding window with respect to $\tau$~(Eqn.~\eqref{eq:sliding_window}), we draw one covariance matrix triplet $k$ from $n_\text{cov}$ possible triplets and update its blocks with the running moment, i.e.,

\begin{align}
&\left(\mathbf{C}_{00,k},\mathbf{C}_{01,k},\mathbf{C}_{11,k} \right) = \\ \nonumber
&\frac{1}{N-\tau}\left(\sum_{i\in I_k}\mathbf{B}_{i}\mathbf{B}_{i}^{\top},\sum_{i\in I_k}\mathbf{B}_{i}\mathbf{B}_{i}^{\prime \top},\sum_{i\in I_k}\mathbf{B}_{i}^{\prime}\mathbf{B}_{i}^{\prime \top}\right),
\end{align}

\noindent{}where $N$ denotes is the total number of time steps contained in all blocks, and $I =$ $\{ 1, \dots, n_\text{cov} \} =$ $\dot{\bigcup} I_k$.

To compute a cross-validated score, we split the covariance matrices into disjoint training and test sets with indices $I_{\text{train}}\dot\cup I_{\text{test}}=\{ 1, \dots, n_\text{cov} \}$ according to a $k$-fold strategy in which we subdivide the set into $k$ groups and choose one of the groups as test set.
We then aggregate the covariance matrices,

\begin{align}
\mathbf{C}_{00}^{\text{test}}&=w\sum_{k\in I_{\text{test}}}\mathbf{C}_{00,k}, \\
\mathbf{C}_{01}^{\text{test}}&=w\sum_{k\in I_{\text{test}}}\mathbf{C}_{01,k}, \\
\mathbf{C}_{11}^{\text{test}}&=w\sum_{k\in I_{\text{test}}}\mathbf{C}_{11,k},
\end{align}

\noindent{}with an appropriate weight $w$, according to the number of samples in the relevant set. Then we calculate the VAMP-$2$ score for both the training and test data sets, yielding scores $CV_{j}$, $j=1,\ldots,k$. Utilizing the VAMP-$r$ score from Eqn.~\eqref{eqn:vamp-r} and letting $r=2$, one can obtain a CV score. The approximated prediction/consistency error is then $\frac{1}{k}\sum_j CV_{j}$.

Thus, following the notation in Sec.~\ref{sec:theory} and Eqns.~\eqref{eqn:u} and~\eqref{eqn:v}, we can summarize the calculation of the cross-validated VAMP-$2$ score as follows, where $||\cdot ||_F$ is the Frobenius norm~\footnote{This is because the Frobenius norm is equivalent to the $r$-Schatten norm for $r=2$; see Ref.~\onlinecite{wu2017vamp}, Sec.~3.2.}, and $\mathbf{U}$ and $\mathbf{V}$ are calculated from the training set:

\begin{align}
A &\equiv \left(\mathbf{U}^{T}\mathbf{C}_{00}^{\text{test}}\mathbf{U}\right)^{-\frac{1}{2}} \nonumber \\ 
B &\equiv \left(\mathbf{U}^{T} \mathbf{C}_{01}^{\text{test}}\mathbf{V}\right) \nonumber \\
C &\equiv \left(\mathbf{V}^{T}\mathbf{C}_{11}^{\text{test}}\mathbf{V} \right)^{-\frac{1}{2}}   \nonumber \\
CV(\bar{\mathbf{K}}(\tau)^{\text{train}}\mid \mathbf{C}_{\cdot,\cdot}^{\text{test}}) &= \| ABC \|_F^r.
\end{align}

\noindent{}We note that the construction of an MSM is not required to obtain the score.

\subsection{Markov state models}
To verify the previously computed VAMP scores on the input feature space, we build MSMs for Homeodomain, Protein G, and WW domain. We expect to see a similar relative VAMP score for the MSMs constructed from different feature sets, which is related to the timescales of the scored dynamical processes.

After obtaining the VAMP singular vectors as described above, the MSM estimation pipeline can be described as follows:~prior to clustering, we project onto the first five right singular vectors of the VAMP basis, analogously to the established TICA method~\cite{perez2013identification}. We then cluster this space into a set of cluster centers via $k$-means, and estimate a maximum-likelihood, reversible MSM on this discretization. For each MSM we perform a block splitting of the discrete trajectories to perform a cross-validated VAMP-2 scoring on the MSM as described in Sec.~\ref{sec:theory} and Ref.~\onlinecite{wu2017vamp}. 
For further analysis we switch from the maximum-likelihood estimate to a Bayesian approach \cite{trendelkamp2015estimation}, in order to compute errors. We evaluate the timescales and some representative structures for the slowest processes.
Finally, we want to ensure that the models have predictive power under the Markovian approximation, so we evaluate the implied timescales and conduct a Chapman-Kolmogorov test (recall Eqns.~\eqref{eqn:implied_timescales} and \eqref{eqn:ck}).
This procedure is outlined in Fig.~\ref{fig:procedure}.

\section{Results}

In Fig.~\ref{fig:vamp2_all_sys_k5} we show VAMP-2 scores for the five slowest processes for all twelve fast-folding proteins at a lag time of $100$ ns computed in the input space as described in the previous section.
For the systems CLN025, Trp-cage, BBA, Villin, Protein B, and Homeodomain, the combination of flexible torsions and the residue contact distance transformation $e^{-d}$ (feature definition~\ref{def:feat_combined}) yields the best overall score (henceforth referred to as the ``combined'' feature set).
For NTL9, BBL, Protein G, $\alpha$3d, and $\lambda$-repressor, a distance-based feature is superior.
The binary contact-based feature scores increase with the magnitude of the cutoff, but are observed or expected to decrease sharply when the cutoff becomes too large to be meaningful.
For all systems, the per-residue SASA feature yields the lowest score.
The results when scoring the ten slowest processes---instead of the five slowest---are comparable; nine of twelve optimal feature sets are the same, and the systems with differing optimal features between five and ten slow processes have multiple comparably well-performing feature sets.
The ten-process analogue of Fig.~\ref{fig:vamp2_all_sys_k5} is shown in Fig.~S1 in the supplementary material.

For all twelve systems, we also analyzed the correlation of each of the five slowest processes (i.e., the five scored VAMP singular components) with the RMSD to the folded structure using the Spearman rank correlation coefficient~\cite{spearman1904proof}.
Ten systems---all except BBL and NTL9---feature a slow process that is well-correlated (0.58 correlation or greater) with folding.
For Villin, the process most correlated with folding is the second-slowest process, and for the other nine folding is the slowest process.
BBL and NTL9 do not feature any slow processes that are well-correlated with folding (maximum correlations of $0.30$ and $0.20$, respectively, for any of the five slowest processes).
Despite the length of these folding datasets, it is therefore important to validate models created for these systems from the simulation data.
The full results for all correlations are presented in Table~S1 in the supplementary material.

\begin{figure*} 
\centering
\includegraphics[width=1\textwidth]{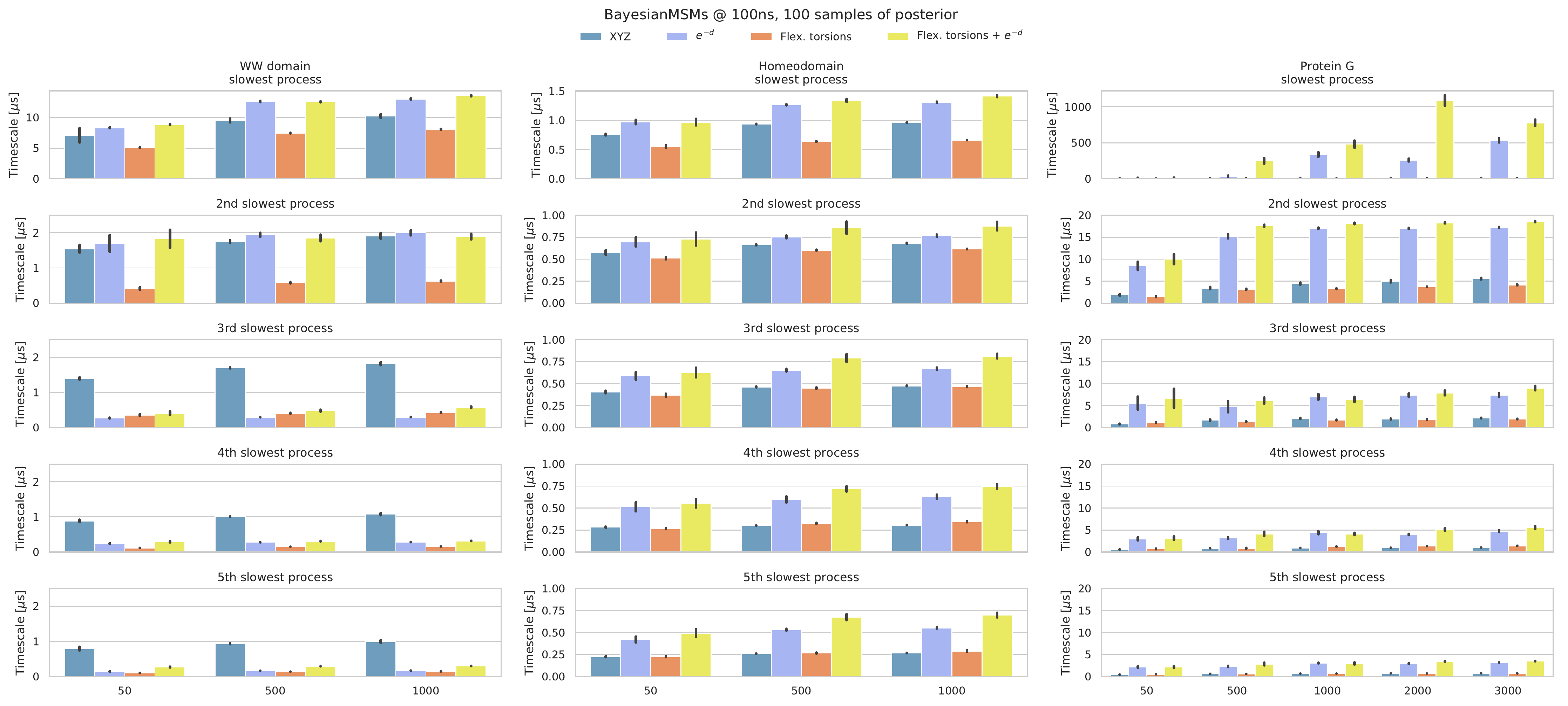}
\caption{Bayesian MSM~\cite{trendelkamp2015estimation} timescales of the five slowest processes in WW Domain, Homeodomain and Protein G for four feature sets and different numbers of cluster centers for 100 sampled transition matrices. The SASA timescales were computed but are too fast to visualize.}
\label{fig:ts_3}
\end{figure*}

To visualize results for a couple of systems, we consider WW domain, Homeodomain, and Protein G.
In Fig.~\ref{fig:ts_3}, we show the MSM timescales for a set of three different $k$-means clusterings---namely $50$, $500$, and $1000$ cluster centers---for each of the three systems.
For all three systems, the first and second slowest processes are nearly always best represented by the combined feature set.
For Homeodomain and Protein G, this trend continues through the fifth slowest process, whereas for WW domain, the aligned Cartesian coordinates (XYZ) feature achieves much longer timescales for the third through fifth slowest processes (further discussion to follow).
All three systems show a slight increase in the process timescales for finer discretizations, which is expected in the absence of cross-validation~\cite{sarich2010approximation,mcgibbon2015variational}. The timescales of the second and following slowest processes show less variance with respect to the number of cluster centers.

\begin{figure*}[tb!]
\centering
\includegraphics[width=1\textwidth]{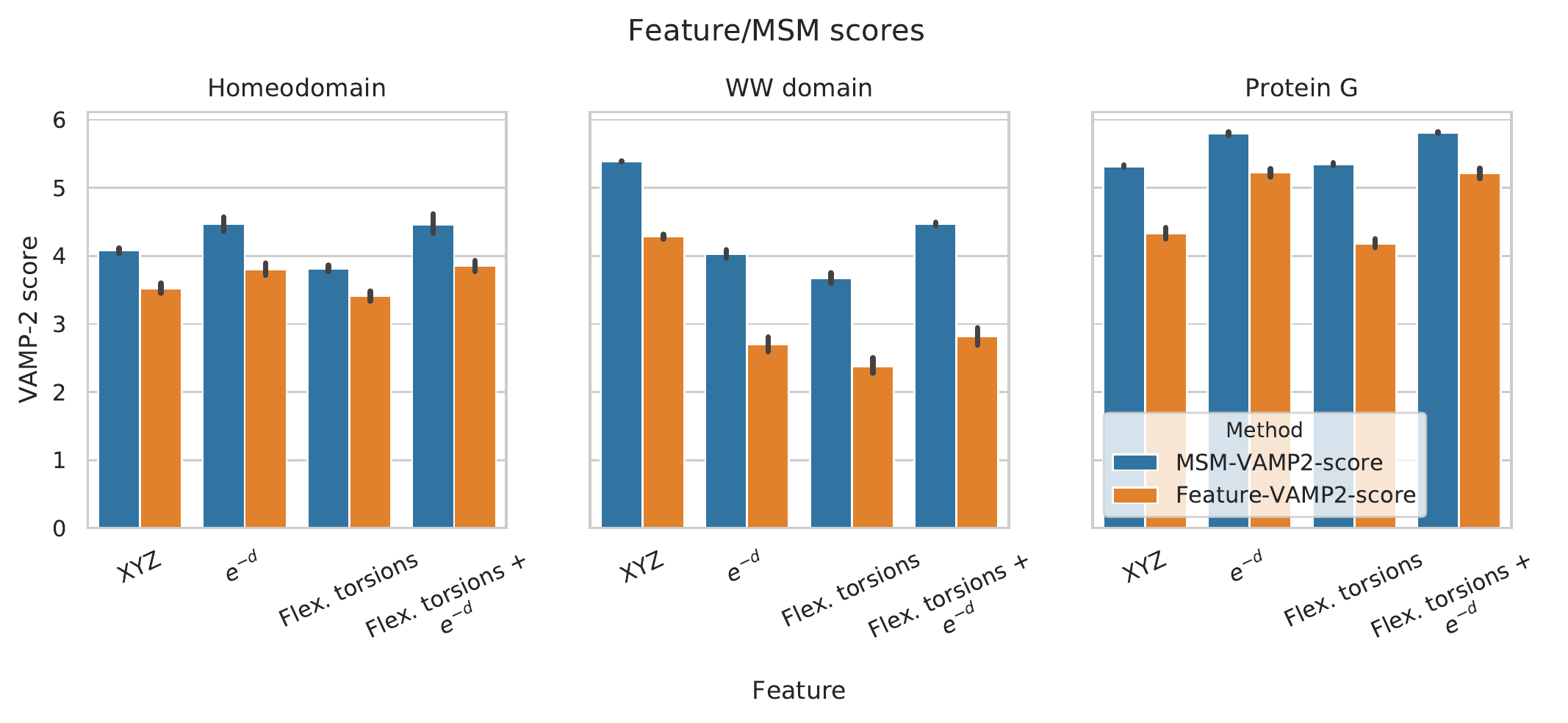}
\caption{Comparison of maximum-likelihood MSM and feature VAMP scores for Homeodomain, WW-domain, and Protein G for four representative feature sets. The MSMs were constructed using 1000 microstates. Both sets of error bars are obtained from 50 cross-validation splits. The feature VAMP columns are reproduced from Fig.~\ref{fig:vamp2_all_sys_k5}.}
\label{fig:feat_vs_vamp_score}
\end{figure*}

\begin{figure}[b!]
\includegraphics[width=0.45\textwidth]{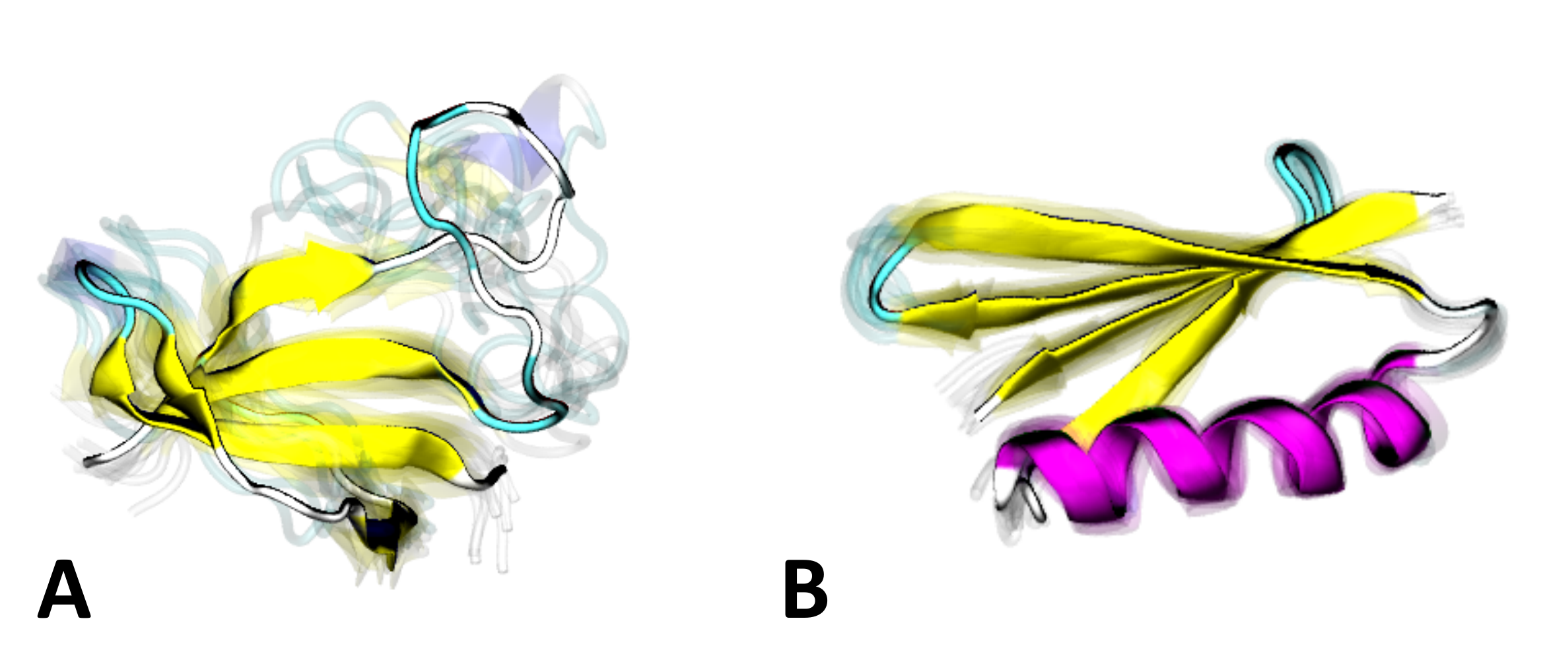}
\caption{Representative structures of Protein G for the slowest process (transition $\mathbf{A} \leftrightarrow \mathbf{B}$) captured by a BayesianMSM~\cite{trendelkamp2015estimation} with a lag time of $100$~ns, built upon 2000 microstates and combined $e^{-d}$ transformed distances and flexible torsions feature.
Amino acid residues are colored according to whether they are belong to an $\alpha$-helix (magenta), $\beta$-sheet (yellow), turn (cyan), or coil (white), and the structures were visualized with VMD~\cite{vmd}.}
\label{fig:protein_g_slowest_proc_flex_t_expd_2000_states}
\end{figure}

\begin{figure}[b!]
  \centering
  \includegraphics[width=0.45\textwidth]{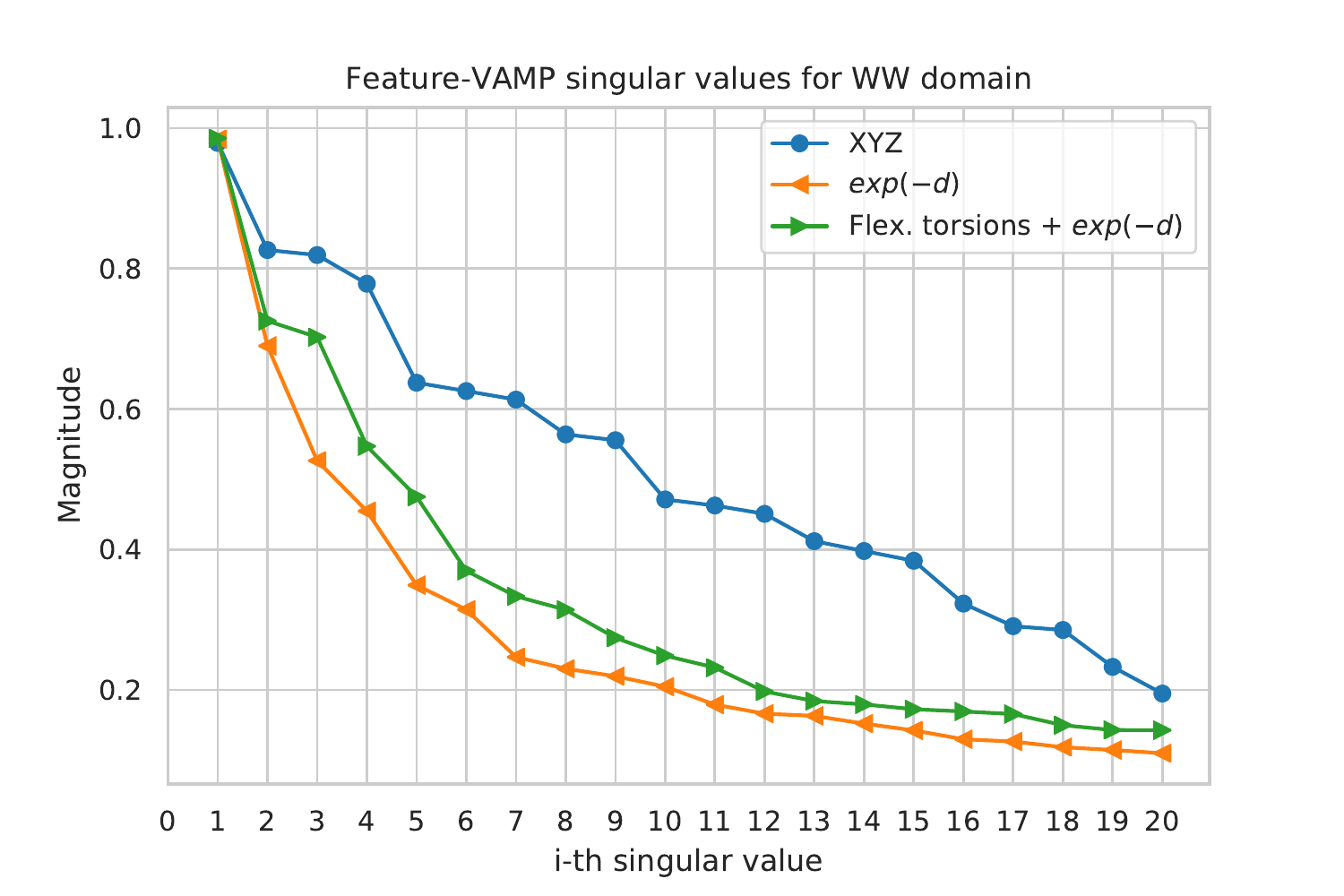}
  \caption{First 20 singular values of VAMP computed on aligned Cartesian coordinates, transformed ($e^{-d}$) distances, and the combination of transformed distances and flexible torsions.}
  \label{fig:cmp_singular_val_ww_domain_xyz_expd}
\end{figure}

Our analysis shows that the torsions feature alone can capture only processes happening on timescales approximately half as long as those processes that can be described after combining torsions with the transformed contact distances.
The $e^{-d}$ transformed contact distance feature alone is relatively effective, but its effectiveness is increased by combining it with flexible torsions.
Aligned Cartesian coordinates generally yield timescales in same regime as the torsion model, with the exception of WW domain, discussed below.
The SASA feature does not capture any slow processes, as we already expected from the low score in the feature space.

In Fig.~\ref{fig:feat_vs_vamp_score}, we also see that the differences in MSM VAMP-$2$ scores for 1000-microstate MSMs built from different features correspond to the trends observed in the previously calculated VAMP-$2$ feature scores.
The VAMP analysis of feature sets can be viewed as pessimistic relative to the full MSM variational analysis (see discussion in Sec.~\ref{sec:discussion}, to follow), since
MSMs increase the variational scores.
However, this is observed to occur across the board, without causing a change in trends. Thus, we observe that, for the goal of choosing the best-performing MSM, we likely could have restricted ourselves to building models from only the best-performing features in the initial analysis.
We note that the construction of an MSM for further analysis should involve investigating many state decompositions (i.e., numbers of microstates, locations of cluster centers) under cross-validation~\cite{mcgibbon2015variational}.
By definition, optimizing the MSM state decomposition will increase the timescales modeled.
However, this must be done under cross-validation, in order to avoid artificial increases in timescales due to overfitting~\cite{mcgibbon2015variational}.

We see that the slowest process in the best-performing Protein G MSMs is extremely slow ($\sim 1$~ms). We visualize the dominant process of a 2000-microstate MSM built from transformed distances in Fig.~\ref{fig:protein_g_slowest_proc_flex_t_expd_2000_states} and see that the process represents conversion between an unfolded structure (\textbf{A}) and the folded structure (\textbf{B}).
The unfolded structure has consistently ordered regions, so the slowest process appears to represent the formation of the $\alpha$-helix and the alignment/prolongation of the $\beta$-sheets.
We note that it has been demonstrated through the use of subsequent simulations that this dataset is undersampled~\cite{schwantes2016markov},
so it is likely that this process is observed only a few times and does not represent folding in general.~\footnote{We also constructed MSMs with cross-validated state decompositions and still observed this timescale.}

\begin{figure}[tb!]
  \centering
  \includegraphics[width=0.45\textwidth]{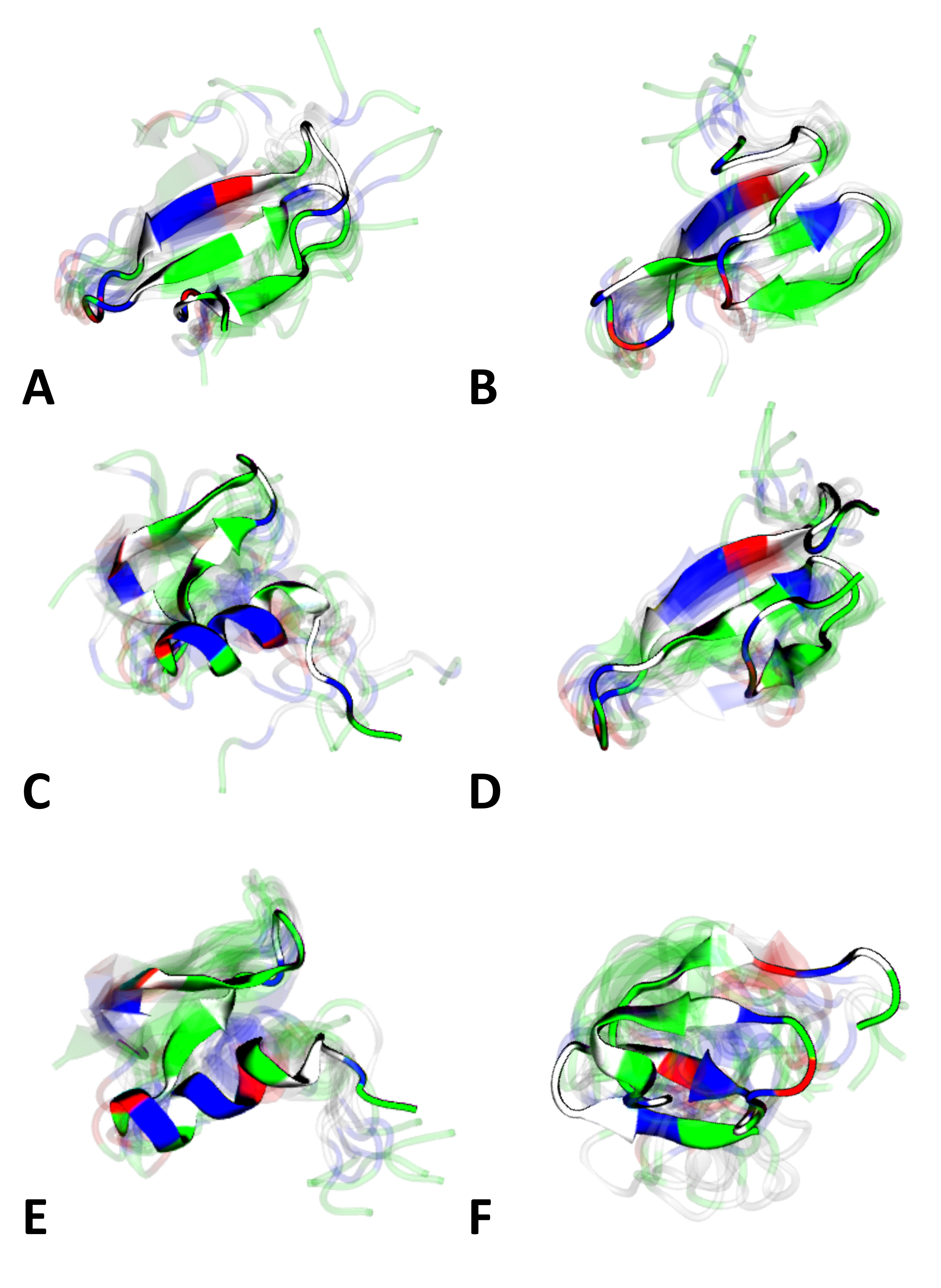}
  \caption{Representative structures for the three slowest processes in WW domain (slowest process $\mathbf{A} \leftrightarrow \mathbf{B}$, second slowest process $\mathbf{C} \leftrightarrow \mathbf{D}$, and third slowest process $\mathbf{E} \leftrightarrow \mathbf{F}$)
  in WW domain extracted from a BayesianMSM with $1000$ states for the combined $e^{-d}$ transformed distances and flexible torsions feature.
  Notably, the slowest process $\mathbf{A} \leftrightarrow \mathbf{B}$ identifies a register shift in the alignment of the $\beta$-sheets.
  Amino acid residues are colored according to whether they are basic (blue), acidic (red), polar (green), and nonpolar (white), and the structures were visualized with VMD~\cite{vmd}.
  }
  \label{fig:gtt_slowest_proc_flex_torsions+expd_and_xyz}
\end{figure}

For all systems except WW domain, the aligned Cartesian coordinates (XYZ) demonstrate poor performance relative to distance-based features.
The WW domain, however, shows the largest score for aligned Cartesian coordinates, which is significantly higher than all the other feature sets. 
Because we only observe this result in one system, we take a closer look and plot the first twenty singular values for three feature sets in Fig.~\ref{fig:cmp_singular_val_ww_domain_xyz_expd}.
We can additionally see from Fig.~\ref{fig:ts_3} that distance-based features exhibit longer timescales for the first and second slowest processes, but very long timescales for the third through fifth processes, which leads to the higher overall VAMP scores for XYZ in Figs.~\ref{fig:vamp2_all_sys_k5} and~\ref{fig:feat_vs_vamp_score}.

We further investigate MSMs constructed from the WW domain data for the combined and XYZ feature sets.
We find that the 1000-microstate MSM built from the combined feature identifies meaningful processes in WW domain folding (Fig.~\ref{fig:gtt_slowest_proc_flex_torsions+expd_and_xyz}).
Particularly, the slowest process identifies a register shift in the alignment of two of the $\beta$-sheets.
Such register shifts have been previously reported for WW domain and other $\beta$-sheet systems in this dataset~\cite{beauchamp2012simple, wan2016maximum}.
In contrast, the 1000-microstate MSM built from aligned Cartesian coordinate features does not yield informative processes.
While the slowest XYZ MSM process shows a transition between the folded state and the denatured ensemble, the second and third slowest processes are overfit to transitions within the folded state (see Fig.~S5 in the supplementary material).

While overfitting may be avoided through the construction of cross-validated MSMs~\cite{mcgibbon2015variational}, the discrepancies between the dynamical processes revealed by the combined feature MSM and the XYZ MSM are already obvious from the analysis presented.
Notably, the XYZ MSM does not resolve the register shift in its three slowest processes.
Thus, despite the apparently superior performance of the aligned Cartesian coordinate feature for WW domain according to its VAMP score only, further investigation shows that it is less useful for analysis.
Therefore, we do not recommend the consideration of aligned Cartesian coordinates as effective features for this type of analysis.

\section{Discussion} \label{sec:discussion}

The introduction of variational principles into MD analyses over the past five years has provided the tools to optimize the approximation quality of not only MSMs, but also the features used to construct reversible or nonreversible dynamical models from features, either as models themselves or as a step toward eventual MSM or Koopman model construction.
With these tools, we have presented a rigorous algorithm to optimize the feature selection step in traditional MSM construction. While subsequent steps such as TICA, cluster discretization, and transition matrix approximation have been relatively optimized, feature selection has remained a challenge for constructing optimal models. 

The method presented in this study demonstrates how recent algorithmic advances enable objective evaluation given a set of feature options.
By optimizing features before building MSMs, we can evaluate their performance directly and more efficiently, instead of in combination with whatever other modeling choices are needed to obtain the MSM. 
We showed that selecting features using the VAMP differentiates them according to the approximation quality of the Koopman matrix obtained from those features. Subsequently constructing MSMs (i.e., from the right singular VAMP vectors) and evaluating their timescales maintains and verifies the ranking from performing VAMP scoring on the features alone.
We expect that the practical use of this algorithm will be to quickly eliminate poorly performing features sets, so that further optimization can be performed on subsequent parameter choices using the best features. Our results show that contact distances transformed by $e^{-d}$ combined with flexible torsions is a consistently well-performing feature set for the folding of small globular proteins, and is likely to perform well for a broader class of conformational transitions in biomolecules. Additional and more specific optimization of features for a class of molecular systems can be done by following the methodology described here.

As shown, constructing MSMs tends to improve model approximation quality due to the state discretization (e.g., transition regions can be finely discretized, which is known to produce better models~\cite{sarich2010approximation}). The discretization process involved in MSM construction will affect features differently depending on the extent to which they are already discretized. For example, nontransformed contact distances will be finely discretized along the coordinate according to the MSM states, whereas a binary contact map cannot be further discretized via an MSM state decomposition. Thus, a continuous coordinate may demonstrate substantial improvement according to a variational score upon MSM construction, whereas an already-discretized coordinate is not expected to improve with MSM construction.

Several studies have already been performed on the choice of MSM hyperparameters following featurization.
A systematic investigation of variationally optimized MSM construction on the same dataset~\cite{husic2016optimized} showed that the use of $10$ or fewer TICs improved model quality, as well as reinforced several other examples of the superior performance of $k$-means clustering for the state decomposition~\cite{scherer2015pyemma,husic2017ward,zhou2017bridging,razavi2017markov}.
Notably, \citet{husic2016optimized} did not report any consistent trends in featurization; however, the research was performed on the protein $\alpha$-carbon backbones only, so the results are not directly comparable to the ones obtained here, which were performed with all protein atoms.
Notably, due to the absence of other heavy atoms, the \citet{husic2016optimized} study did not investigate featurization with flexible torsions, so the relatively poor performance of flexible torsions in this work cannot be compared with the previous analysis.
However, since $\phi$ and $\psi$ torsional angles tend to encode secondary structure (whereas contacts are more closely related to tertiary structure) it is unsurprising that using torsions alone lead to inferior models, but also unsurprising that their inclusion in a combined feature with contact information generally improves models when compared to representations with contact information only.

Although the types of features investigated were derived from the protein backbones only, \citet{husic2016optimized} show that the choice of feature limits the effectiveness of the model in resolving the slow dynamical processes, despite any dimensionality reduction and clustering that follows.
However, the authors do not suggest how this important issue might be systematically approached.
Here, we present a suitable and rigorous algorithm for exactly this problem.
To build an optimal MSM for a given system, the method derived in the present work and its results are just the first step, and provide a systematic starting point that can be followed by incorporating the findings of already-established investigations and the trends they show. 

The VAMP approach is based the fact that slow dynamics determine the global kinetics of a system. However, it is well-known that sometimes the slowest processes are irrelevant to the dynamics under study, such as a rare dihedral flip~\cite{banushkina2015nonparametric}.
Therefore, if the slow processes are not the ones of interest, then the optimization of the global kinetics diverges from the modeling of the processes of interest.
However, it is still the former that must be optimized in order to create the model.
Currently, there is no systematic way to model selected kinetics with respect to a chosen observable (i.e., a kinetic analogue of the functional mode analysis presented in Ref.~\onlinecite{krivobokova2012partial}), and such an algorithm is left to future work.

Whereas in this study we apply the method to datasets that are approximately in thermodynamic equilibrium, we emphasize that the procedure detailed in this section and Sec.~\ref{sec:theory} is general to equilibrium or nonequilibrium datasets.
Unlike the VAC and TICA, VAMP avoids the bias associated with the symmetrization of the dynamics, which can be a problem for models of reversible systems with rare events~\cite{wu2017variational}.
Furthermore, one does not need to assess whether a reversible or nonreversible approximation is appropriate for the data at hand.
For a similar application to nonequilibrium data, we refer the reader to Ref.~\onlinecite{paul2018identification}.

Since the first applications of the MSM framework to the analysis of biomolecules simulated with MD, the decision of how to transform the raw Cartesian coordinates from a simulation dataset into well-performing features has been an omnipresent challenge. While the categorical aspect of feature selection prohibits pure automation (except, perhaps, in the case of treatment with neural networks~\cite{mardt2018vampnets}), the method presented in this work enables a direct evaluation of features from which predictive models can be more efficiently built.

\section*{Supplementary material}

See supplementary material for additional results on VAMP-optimized models and MSM validation.

\section*{Acknowledgements}

We are grateful to Tim Hempel, Nuria Plattner, Andreas Mardt, and Simon Olsson for helpful discussions.
We thank D.~E.~Shaw Research for providing simulation datasets.
Funding is acknowledged from 760 European Commission (ERC CoG 772230 ``ScaleCell''), 761 Deutsche Forschungsgemeinschaft (SFB 1114 Projects 762 C03 and A04, SFB 740 Project D07, NO825/2-2), and 763 the MATH+ cluster (Project EF1-2).

\bibliography{refs}

\end{document}